%% file: main.tex
\pdfoutput=1
\documentclass[sigconf, screen, nonacm=true]{acmart}

\usepackage{svg}
\usepackage{subcaption} 
\usepackage{tcolorbox}
\usepackage{multirow}

\newcommand{\ignore}[1]{}

\newcommand{\Rbb}{\mathbb{R}}
\newcommand{\vectorize}{\text{vec}}

\AtBeginDocument{%
  \providecommand\BibTeX{{%
    \normalfont B\kern-0.5em{\scshape i\kern-0.25em b}\kern-0.8em\TeX}}}

\setcopyright{acmcopyright}

\copyrightyear{2023}
\acmYear{2023}
\setcopyright{acmlicensed}\acmConference[ISCA '23]{Proceedings of the 50th Annual International Symposium on Computer Architecture}{June 17--21, 2023}{Orlando, FL, USA}
\acmBooktitle{Proceedings of the 50th Annual International Symposium on Computer Architecture (ISCA '23), June 17--21, 2023, Orlando, FL, USA}
\acmPrice{15.00}
\acmDOI{10.1145/3579371.3589044}
\acmISBN{979-8-4007-0095-8/23/06}

\begin{document}

\title{Enabling High Performance Debugging for Variational Quantum Algorithms using Compressed Sensing}


\author{Kun Liu}
\authornote{Both authors contributed equally to this research.}
\orcid{0009-0002-9038-8733}
\affiliation{%
  \institution{Carnegie Mellon University}
  \city{Pittsburgh}
  \state{PA}
  \country{USA}
  \postcode{15213}
}
\email{liukun@cmu.edu}

\author{Tianyi Hao}
\authornotemark[1]
\orcid{0000-0003-4074-4971}
\affiliation{%
  \institution{University of Wisconsin-Madison}
  \city{Madison}
  \state{WI}
  \country{USA}}
\email{tianyi.hao@wisc.edu}

\author{Swamit Tannu}
\orcid{0000-0003-4479-7413}
\affiliation{%
  \institution{University of Wisconsin-Madison}
  \city{Madison}
  \state{WI}
  \country{USA}
}
\email{swamit@cs.wisc.edu}

\renewcommand{\shortauthors}{Liu and Hao, et al.}

\input{sections/0abstract}




\maketitle

\input{sections/1intro} 
\input{sections/2background} 
\input{sections/3problem} 
\input{sections/4OSCAR} 
\input{sections/5OSCAR_D}
\input{sections/6ucase-1} 
\input{sections/6ucase-4}
\input{sections/6ucase-3}
\input{sections/8RelatedWorks} 
\input{sections/9Conclusion}
\appendix
\input{sections/10Appendix}

\bibliographystyle{ACM-Reference-Format}
\balance
\bibliography{Bibs/refs}

\end{document}

%% file: sections/0abstract.tex
\begin{abstract}
Variational quantum algorithms (VQAs) can potentially solve practical problems using contemporary Noisy Intermediate Scale Quantum (NISQ) computers. VQAs find near-optimal solutions in the presence of qubit errors by classically optimizing a loss function computed by parameterized quantum circuits. However, developing and testing VQAs is challenging due to the limited availability of quantum hardware, their high error rates, and the significant overhead of classical simulations. Furthermore, VQA researchers must pick the right initialization for circuit parameters, utilize suitable classical optimizer configurations, and deploy appropriate error mitigation methods. Unfortunately, these tasks are done in an ad-hoc manner today, as there are no software tools to configure and tune the VQA hyperparameters.  

In this paper, we present OSCAR (cOmpressed Sensing based Cost lAndscape Reconstruction) to help configure: 1) correct initialization, 2) noise mitigation techniques, and 3) classical optimizers to maximize the quality of the solution on NISQ hardware. OSCAR enables efficient debugging and performance tuning by providing users with the loss function landscape without running thousands of quantum circuits as required by the grid search. Using OSCAR, we can accurately reconstruct the complete cost landscape with up to 100X speedup. Furthermore, OSCAR can compute an optimizer function query in an instant by interpolating a computed landscape, thus enabling the trial run of a VQA configuration with considerably reduced overhead.
\end{abstract}

%% file: sections/1intro.tex
\section{Introduction}
Rapid progress in quantum computing has brought together a broad community of researchers who want to leverage quantum computers to solve problems in areas ranging from high-energy physics to finance. Using a quantum computer to solve scientific and commercial problems will require full-stack solutions and system-level abstractions that can address a diverse group of end users and allow them to access quantum computers seamlessly. This paper focuses on building methods to help debug, tune, and benchmark near-term quantum algorithms. 

Variational Quantum Algorithm (VQA)
~\cite{peruzzo_variational_2014}
is one of the most promising categories of quantum algorithms that can utilize existing quantum computers with hundreds of noisy qubits to solve hard problems in optimization, chemistry, and machine learning. The paradigm of VQA is similar to that of Machine Learning (ML). In ML, we train a model by searching for the optimal parameters of the model on the loss landscape. In VQA, the difference is that the model is a variational quantum circuit (VQC), a quantum circuit with parameters. The VQC is executed on the quantum computer to produce the probability distribution of solutions, which is then used to compute the solutions' average cost. Just like the ML training, the parameters of VQC are tuned using an optimizer running on the conventional computer to maximize or minimize the average cost. The circuit parameters form a continuous parameter space, and as we vary the parameters of VQC, we obtain a cost hypersurface in the parameter space. Like in ML, we also denote it as `landscape'.

Copying data, one of the most common computing primitives, is not available on quantum computers, as the laws of quantum mechanics prohibit the creation of exact copies of quantum information. Moreover, upon reading the qubit, we can only extract partial information about its state. This makes debugging and tuning quantum programs challenging as we can not ``single step'' through the program to probe the intermediate state due to probabilistic qubit readout, nor can we create copies of the data to analyze the intermediate state. In addition to the fundamental constraints, there are engineering challenges. For example, all quantum hardware platforms are extremely noisy, and running programs on error-prone quantum hardware often produces incorrect output. As a result, when we observe an undesired output on a quantum computer, it is hard to differentiate a software bug from a hardware error.

VQAs add to this challenge as they rely on optimizers to search and coverage on the optimal circuit parameters that maximize the quality of the solution. If optimizers are not configured correctly or if we pick an initial point that is close to a local minimum, the final VQA output can be suboptimal, producing low-quality solutions.   Moreover, to improve the efficacy of  VQAs, researchers have developed noise mitigation techniques that execute additional runs to offset or filter the effect of noise. Although well-intentioned and theoretically proven, noise mitigation does work under a narrow set of conditions and can require precise configuration.       

When running VQA on a near-term quantum computer, all components of the system must work in tandem to get high-quality solutions.   For example, the programmer describes the parametric circuit (or ansatz), picks the initial point, configures the optimizer, and configures the noise mitigation technique. To enable the optimal solution on VQA, all previously described steps need to be performed accurately. Any misconfiguration can result in an undesired system behavior or a bug. Unfortunately,  discovering these scenarios can be highly challenging due to:
(1) The inability to trace intermediate states of quantum programs~\cite{huang_statistical_2019,li_projection-based_2020,liu_quantum_2020}. (2) Noisy hardware can further muddle the output~\cite{murali_noise-adaptive_2019,tannu2019not}. (3) Several tightly integrated non-deterministic components, such as optimizer and noise mitigation schemes.

To debug the VQAs, we must untangle and check components in isolation. We observe that using an optimizer adds non-determinism. Furthermore, the path traversed by the optimizer is highly sensitive to the landscape.   
One of the most straightforward ways to reduce non-determinism is by disengaging the optimizer from the overall VQA workflow. Instead of traversing the landscape guided by the optimizer, we propose to leverage an entire landscape. With a full landscape, we could calculate the variance of gradient and probe directly into barren plateaus, check the quality of initial points and convergence of optimization, measure the performance of different quantum error mitigation methods, visualize the landscape, and reveal deeper insights.

However, generating a full landscape by using a grid search, i.e., uniformly sampling on each dimension of the landscape, is extremely challenging due to the high cost of running VQA circuits. In real experiments, thousands of shots are needed for the near-term quantum device to compute a single point on the landscape. Generally, for each point on the landscape, we derive it by running the quantum circuit number-of-shots (usually over 1,024) many times and measuring these quantum states.




We use this insight of sparsity and propose {\em cOmpressed Sensing based Cost lAndscape Reconstruction (OSCAR)}, which leverages the idea of compressed sensing (CS)~\cite{candes_stable_2006,donoho_compressed_2006} to efficiently reconstruct the full landscape by using only a small fraction of the experiments compared to performing a full grid search. CS can enable full reconstruction by leveraging the sparsity of signals in the frequency domain. We discover that optimization landscapes for the majority of VQAs are sparse in the frequency domain, and these landscapes can be faithfully reconstructed by using standard mathematical tools developed by the compressed sensing community. Based on the analysis of the landscape given by CS, we could choose initial points, optimizers, and other settings to achieve better performance. With few points on the landscape, we could reconstruct the whole landscape with reasonable loss of accuracy. We evaluate our technique on simulated landscapes and landscapes generated on Google and IBM superconducting qubit quantum processor.

To reconstruct a full landscape, we sample parameter values randomly and uniformly from the entire parameter space and then run VQA circuits to generate the expected value of the cost functions corresponding to the randomly selected parameters. Using these expected values, we execute a compressed sensing routine that reconstructs a landscape. Our experiments show that in less than a few thousand experiments,  we can reconstruct an entire landscape with minuscule error.   Using a simulator, we test our design in an ideal setting and with depolarizing noise. Furthermore, we use landscape data collected on Google's quantum computers to show that even on real hardware with complex noise sources, we can execute OSCAR to generate an accurate cost landscape significantly faster. 

The idea of using compressed sensing to reconstruct a full landscape from a few random samples is extremely powerful because it detaches the classical optimization routine from Variation Quantum Algorithms. During the debugging process, it reduces the complexity and non-determinism added by the optimization routine, but even more importantly, it enables parallelism as the calculation of expected values is no longer serialized due to the optimizer. In a typical VQA workflow, we run a quantum circuit, compute the expected cost value and use it to determine the next set of circuit parameters using an optimizer. This model enforces serial execution. With OSCAR, we can pick random circuit parameters and run circuits in parallel on multiple quantum computers to further accelerate reconstruction to enable fast debugging. However, using OSCAR as it is in parallel mode is challenging due to variability in error rates. Different quantum hardware platforms have different noise profiles. Even on the same quantum computer, a subset of qubits exhibits large variations in error rates~\cite{tannu2019not}. If we collect samples on devices with different noise-level, the reconstructed landscape would be an "artificial" mixture of original landscapes, masking hardware-specific effects. We propose a noise-compensation algorithm to minimize this mixing to retain the device-specific effects. We transform the expected values obtained on a quantum computer with one noise configuration to another machine with a different noise configuration using linear regression to preserve the impact of noise on the reconstructed landscape. Furthermore, we propose eager reconstruction, wherein OSCAR can fully leverage the speedup enabled by parallel execution despite high tail latency and queuing delays on the real systems.

In this paper, we showcase three use cases of OSCAR. First, we show that by using OSCAR, we can efficiently navigate a complex process of configuring noise mitigation techniques. Noise mitigation is highly effective on near-term quantum computers. Unfortunately, many noise mitigation techniques require a significant number of additional circuit runs, making them expensive. Furthermore, these mitigation techniques need to be carefully configured as they can affect the problem landscape. For example, we observe that noise mitigation techniques make problem landscapes sharp, amplifying the gradients but also they add jaggedness. This can negatively affect the optimization process. We show that by using OSCAR. We can use OSCAR to reconstruct a landscape that preserves certain local traits using only a 10\% fraction of samples and give users the ability to benchmark and test the noise mitigation method without running a substantial number of experiments. OSCAR's second and third use cases focus on selecting and configuring the optimizer. A reconstructed landscape gives us a large number of grid points in the parameter space. We can make the grid dense by using interpolation and running an optimizer on the reconstructed landscape. In this case, we simply query the reconstructed landscape instead of running an actual circuit for specific parameters. We show that by optimizer running with default VQA workflow and on a landscape reconstructed by OSCAR, converge on points very close to each other. This gives us the ability to test optimizers with different configurations without actually querying quantum computers to faithfully capture the interaction between the optimizers and the problem landscape. Furthermore, in the third use case, we show that the interpolated landscape can assist users in picking a good initial point and the right optimizers.




%% file: sections/2background.tex
\section{Background}


\subsection{Quantum Programming Model}
Quantum computers are domain-specific accelerators that leverage properties of quantum bits (qubits) to solve computationally challenging problems. Quantum programs use sequences of quantum gates that manipulate the collective state of qubits. Today, most quantum computers are programmed using a co-processor model, in which the programmer uses a domain-specific language to describe the quantum program. Then, the program is compiled to an executable, which the host machine offloads to {\em Quantum Processing Unit (QPU)}.
Finally, the program is executed, and output is returned to the host machine.
Unfortunately, quantum hardware is prone to errors, and the execution of quantum programs can produce incorrect results. Researchers are actively developing reliable hardware and  noise-resilient algorithms that we can use in the near term.


\subsection{Variational Quantum Algorithms}

\begin{figure}[t]
\centering
\includegraphics[width=\linewidth]{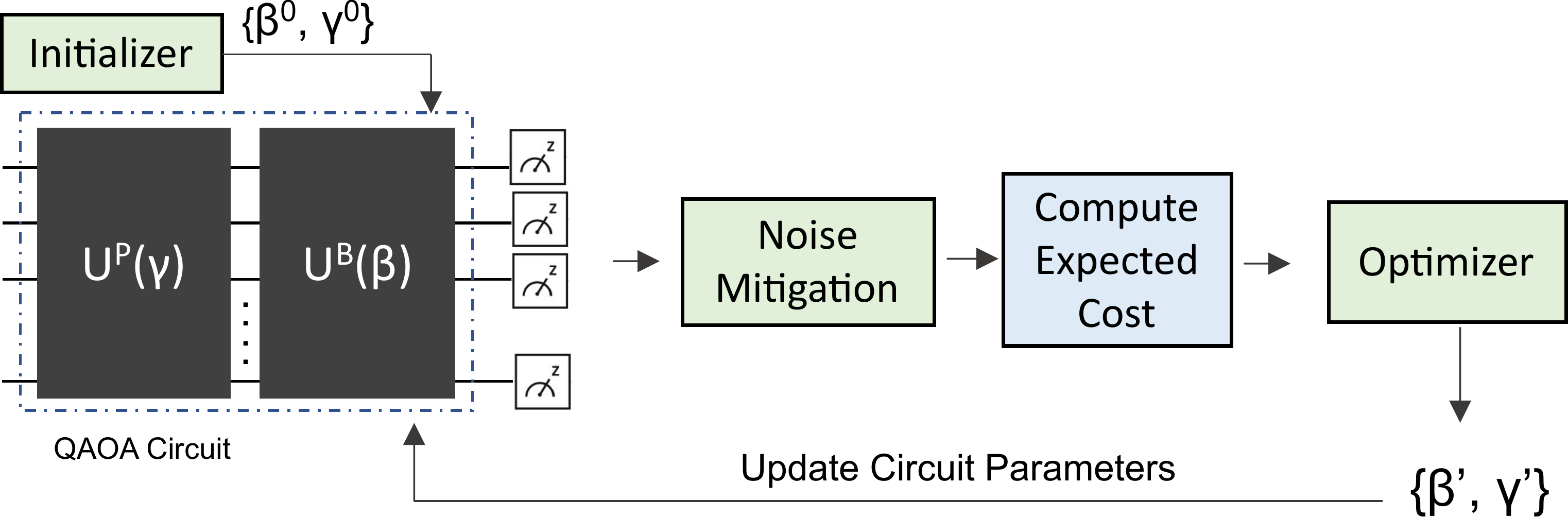}
\caption{Typical workflow used for executing QAOA.}
\vspace{-0.25in}
\label{fig:VQA_intro}
\end{figure}

Variational Quantum Algorithms (VQAs) are a family of quantum algorithms that promise to solve challenging optimization and learning problems using near-term quantum computers. VQAs use a parametric circuit and search iteratively for the circuit parameters that produce high-quality solutions. The Quantum Approximate Optimization Algorithm (QAOA)~\cite{farhi_quantum_2014} is one of the VQAs,
which is widely used to solve combinatorial optimization problems. When solving an optimization problem using QAOA, as illustrated in Figure~\ref{fig:VQA_intro}, we search for circuit parameters $\beta$ and $\gamma$ using a two-step process. First, we initialize $\beta$ and $\gamma$ with the best-known value and execute a quantum circuit several thousands of times. This yields a distribution of solution strings, where each solution has a fixed cost. Our objective is to find the solution string with the lowest cost. Next, we compute the average (or expected) cost corresponding to the output distribution and search for optimal $\beta$ and $\gamma$ using expected cost as the objective function.

\subsection{Noise mitigation with VQA Workflow}

VQA circuits are expected to run on noisy hardware, where the produced output can be incorrect due to qubit errors corrupting the computation. However, most VQAs estimate the expected (or average)  value of a cost function to tune the circuit parameters. Fortunately, the average estimate is resilient to small changes in the output distributions due to errors. However, with current error rates, it is challenging to get accurate results on problems at scale that are useful for practical application. With an average two-qubit gate error rate of 0.5\% on leading commercial hardware platforms, we can run applications with, at most, hundreds of two-qubit gates. To expand the capabilities of noisy quantum computers, we can use noise mitigation methods such as Dynamical Decoupling (DD)~\cite{viola_dynamical_1998,das2021adapt}, Qubit Readout Mitigation (QRM)~\cite{Bravyi2021}, Zero Noise Extrapolation (ZNE)~\cite{li_efficient_2017},  Probabilistic Error Cancellation (PEC)~\cite{temme_error_2017}, and Clifford Data Regression(CDR)~\cite{czarnik_error_2021},categorized as: 

{\noindent\textbf{Shot Frugal Mitigation -}} leverages methods that enhance gate or operational fidelity by augmenting the input circuit without increasing the number of shots. Dynamical decoupling, for example, is implemented by simply inserting a sequence of single-qubit gates to suppress the ZZ-crosstalk~\cite{cai_impact_2021,dicarlo_demonstration_2009,mckay_three-qubit_2019}, while readout mitigation is performed by constructing the inversion matrix and filtering measurement errors using a post-processing step. The shot frugal mitigation methods do not require additional circuit executions.   

{\noindent\textbf{Mitigation with Supplementary Shots -}} In this category, the circuit and circuit configurations are altered at runtime. For example, Zero Noise Extrapolation (ZNE) evaluates the expected value of a cost function with varying noise levels and extrapolates it back to estimate the expected value with zero noise. Several demonstrations at small and large scales have shown encouraging results. 
Unfortunately, these methods require a significantly higher (10x to 100x) number of circuit runs and non-trivial configurations. 

To improve the estimates of expected values on current hardware, the user needs to set up noise mitigation strategies by using trial and error methods. The methods are theoretically sound, but there are noise and circuit-specific factors that require manual intervention, and if the user is not careful, the noise mitigation techniques can do more harm than good. 






%% file: sections/3problem.tex
\section{Debugging and Tuning VQAs}

\subsection{Motivation}

Quantum debugging and analysis tools are essential for developing, deploying, and benchmarking quantum programs. The goal of a software debugger is to help programmers track down incorrect or inconsistent implementation, whereas the analysis tools can help improve the overall reliability and quality of the solution by tuning the quantum programs.  

Building debugging and analysis tools for quantum programs are fundamentally challenging due to (1)  destructive and statistical reads, (2) no-cloning theorem, and (3) uncertainty introduced by noisy operations. Reading a qubit gives only partial information as qubit readout collapses the qubit to `0' or `1' states with a certain probability, and precisely estimating the quantum state can require a significant number of repeated measurements. Furthermore, creating copies of quantum variables is prohibited due to the no-cloning theorem. Both constraints make probing the intermediate state of quantum programs extremely challenging. Huang et al. proposed statistical assertions to navigate this challenge by using carefully designed statistical tests to check if the intermediate state of the quantum program satisfies specific properties~\cite{huang_statistical_2019}.

Using projective measurements and runtime assertions proposed by Li et al.~\cite{li_projection-based_2020} and Liu et al.~\cite{liu_quantum_2020},  we can reduce the overhead of probing the internal state of a quantum program significantly by leveraging additional ancillary qubits. However, for all circuit and statistical debugging techniques, the programmer needs to know the correct state or some properties of the correct state, and by using quantum and statistical gadgets, one could track inconsistencies in the circuit implementations. These works lay the groundwork for debugging quantum circuits and programs. However, prior work does not address debugging and tuning near-term quantum algorithms, such as VQAs running on noisy quantum computers. 

\subsection{Challenges in Debugging and Tuning VQAs}
Debugging quantum circuits is significantly challenging in the presence of noise,  as noisy operations produce incorrect outcomes, and determining if the undesired output is produced due to a hardware error because of noisy qubits or by incorrect implementation can be very difficult. Furthermore, VQAs utilize optimizer-driven circuit parameter updates, resulting in complex internal states of the program that can not be easily used to drive the debugging process. Unlike quantum kernels such as the Quantum Fourier Transform, which is at the heart of Shor's factoring algorithm~\cite{shor_algorithms_1994}, or Amplitude Amplification, used in Grover's Algorithm~\cite{grover_fast_1996}, VQAs do not exhibit structure that programmers can easily use to verify correctness. Similar to classical deep neural networks, there is no one correct internal state that programmers can use as a reference to debug the program~\cite{lipton2018mythos}. Additionally, optimization inserts a significant amount of non-determinism in the program behavior. All VQAs fundamentally rely on a tightly integrated workflow between the quantum computer and optimizer, where misconfiguration on one component can affect the overall correctness.

The complexity of VQA workflows is expected to increase with the introduction of error mitigation methods that run circuits with different configurations. Many noise mitigation techniques require a significantly large number of circuit runs, and they are challenging to configure. We envision that users developing VQAs will need tools to reason with the effectiveness of noise mitigation techniques to tune current and future quantum algorithms. Especially with the current cloud computing model, with limited system-level access.

\begin{figure}[t]
\centering
\includegraphics[width=0.8\linewidth]{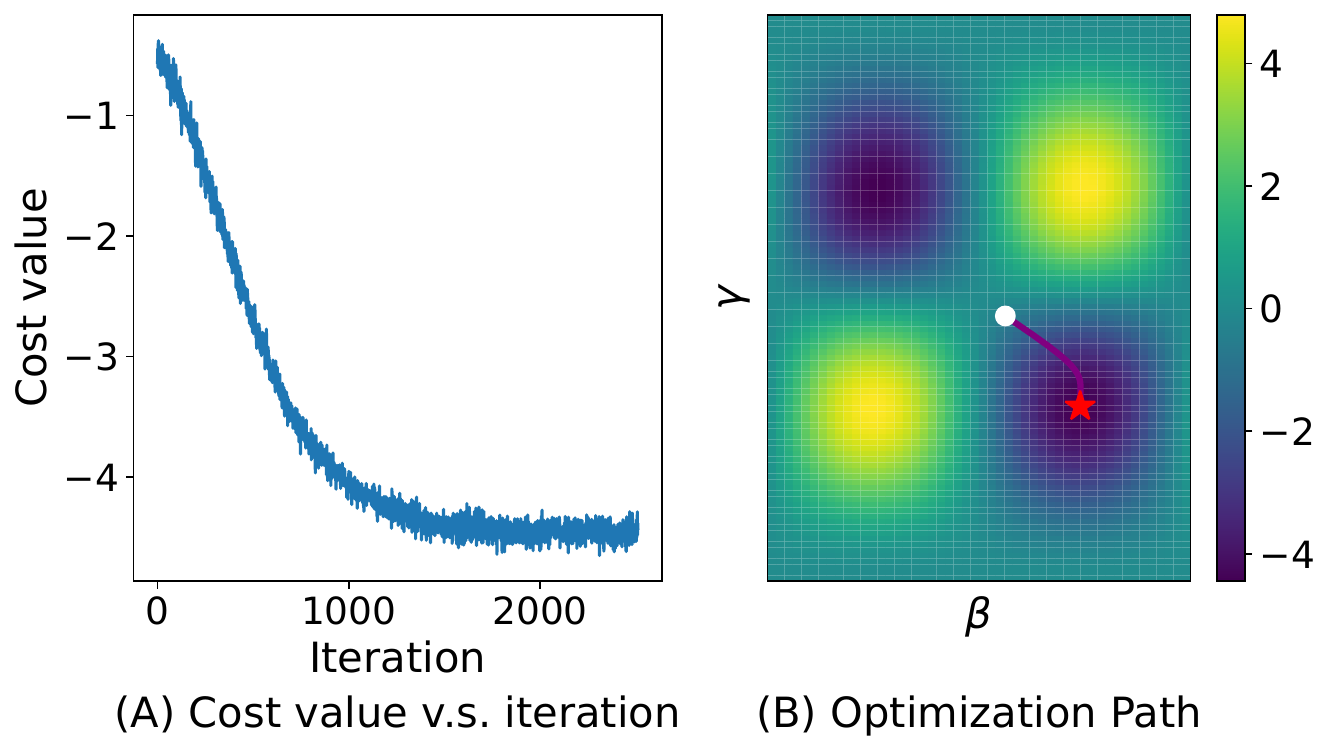}
\caption{(A) Default view provided by the VQA workflow (B) Bird's-eye view with a complete optimization landscape.}
\label{fig:cost_vals_and_opt_path}
\vspace{-0.2in}
\end{figure}

\subsection{How complete landscape can help?}
When debugging VQAs, we want to understand which workflow component is causing the undesired output. Unfortunately, the current software workflows provide a narrow view of the landscape as the optimizer picks the path (a tiny fraction from a cost function landscape) traversed on the landscape. The objective function of the optimizer is designed to maximize the quality of the solution. However, VQA landscapes are riddled with local maxima/minima and barren plateaus~\cite{mcclean_barren_2018,marrero_entanglement_2021,cerezo_cost_2021},  which are further exacerbated due to noisy operations, making it challenging to establish the impact of specific configuration and their effect on the output quality when running an optimizer. We need context to understand how optimizer configuration affects the traced path. For example, assessing the efficacy of design decisions would become significantly easy if we could access the entire landscape from a bird's eye view. Figure~\ref{fig:cost_vals_and_opt_path}, show two points of view, panel  (A) shows optimizer centric view, which users see when they use standard workflow, whereas panel (B) shows the path traced by the optimizer superimposed on the landscape, providing context for the user to understand the dynamic between optimizer, landscape, and noise mitigation. 

For prototyping and as a pedagogical tool, complete landscapes are often generated by doing a grid search, where a large number of $\beta$ and $\gamma$ values are evaluated on a quantum computer or the simulator. However, grid search is extremely slow and expensive, especially for debugging, a highly uncertain process, where the user does not know a priori how many executions they will need to trace the bug.



\begin{figure*}[t]
\centering
\includegraphics[width=0.8\linewidth]{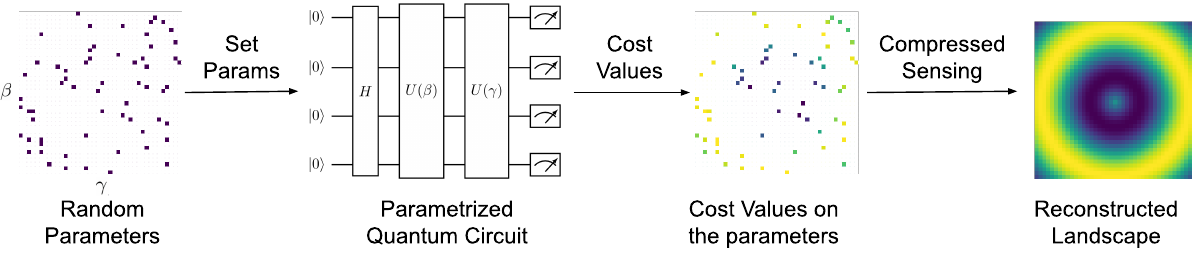}
\caption{Overview of OSCAR illustrating three steps: random parameter sampling, circuit execution, and  reconstruction.}
\vspace{-0.20in}
\label{fig:diagram_oscar}
\end{figure*}

%% file: sections/4OSCAR.tex
\section{Landscape Reconstruction by OSCAR}
We propose {\em cOmpressed Sensing based Cost lAndscape Reconstruction (OSCAR)}. An open-source implementation in Python is available at \href{https://github.com/haoty/OSCAR}{\color{blue}{https://github.com/haoty/OSCAR}}. This section details our design and demonstrates its effectiveness in practice.

\vspace{-0.05in}

\subsection{Overview of OSCAR Workflow}
OSCAR leverages compressive sensing to generate the entire cost landscape using a small number of samples. Our design is comprised of three phases: parameter sampling, circuit execution, and landscape reconstruction, as shown in Figure~\ref{fig:diagram_oscar}.


{\noindent \textbf{Parameter Sampling.}} We randomly select a small number of circuit parameters that are uniformly distributed over the parameter space (shown as $\beta$ and $\gamma$ points in Figure~\ref{fig:diagram_oscar}). The dimension of the optimization landscape scales with the number of independent parameters, which is $2p$ for a QAOA circuit with $p$ layers. When reconstructing high-dimensional landscapes, we perform concatenations to reduce the dimension.

{\noindent \textbf{Circuit Execution.}} We execute the VQA circuit with the sampled circuit parameters for a set number of shots to calculate the expected values of the cost function. 

{\noindent \textbf{Landscape Reconstruction.}} We use the cost function values from the previous phase to reconstruct the landscape using compressed sensing. Compressed sensing (CS)~\cite{candes_stable_2006,donoho_compressed_2006} reconstructs a signal with samples significantly fewer than what Nyquist–Shannon sampling theorem requires. Nyquist-Shannon sampling theorem states that if the sampling rate is two times higher than the highest frequency of a signal, then this signal could be reconstructed perfectly. However, the constraint on the sampling rate is not absolute. We can enable high-quality reconstruction with significantly fewer samples if the signal has a small number of frequency components, that is if the signal is sparse in the frequency domain. We observe that {\em VQA landscapes are sparse in the frequency domain and can be accurately reconstructed using compressed sensing.}

We leverage this insight to generate the entire landscape by running a significantly lower number of circuits compared to the standard grid search. For more mathematical details of CS, please refer to Appendix~\ref{ap:cs_math}.


\begin{table}[h]
\centering
\resizebox{\columnwidth}{!}{%

\begin{tabular}{llll}
\toprule
    Depth & $\beta$ range, \# samples & $\gamma$ range, \# samples & Total \# samples    \\
\midrule
p=1 & $[-\pi/4, \pi/4]$, 50     & $[-\pi/2, \pi/2]$, 100     & $50\times100$ = 5k  \\
p=2 & $[-\pi/8, \pi/8]$, 12     & $[-\pi/4, \pi/4]$, 15      & $12^2 \times 15^2$ = 32k \\
\bottomrule
\end{tabular}%
}
\caption{Grid Definition of QAOA Ansatz.}
\label{tab:sampling_range}
\vspace{-0.4in}
\end{table}

\subsection{Reconstruction Accuracy of OSCAR}
We evaluate the reconstruction accuracy of OSCAR using simulators and data collected on IBM and Google hardware.

\subsubsection{Method}
We use a total of 200 problem instances of QAOA to evaluate the landscape reconstruction accuracy of OSCAR. To generate the ground truth, we perform a dense grid search to obtain the true landscape, by executing 5,000 to 32,000 quantum circuits corresponding to the grid points on the landscape as shown in Table~\ref{tab:sampling_range}. Each circuit is executed using state vector simulation.

\begin{table}[b]
\renewcommand{\arraystretch}{1.8}
\centering
\setlength{\tabcolsep}{0.035cm}
\renewcommand{\arraystretch}{1.2}
\resizebox{\columnwidth}{!}{%
\begin{tabular}{cccccc}
\toprule
Problem      & \#Qubits & \#Parameters & \#Samples & QAOA & Two-local \\
\midrule
3-reg MaxCut & 4      & 8  & 7   & 0.847       & 0.645           \\
3-reg MaxCut & 6      & 6  & 14   & 0.372       & 0.0000001       \\
SK Problem   & 4      & 8  & 7   & 0.847       & 0.765           \\
SK Problem   & 6      & 6  & 14   & 0.372       & 0.057          \\
\bottomrule
\end{tabular}%
}
\caption{
Reconstruction errors for QAOA and Two-local ansatzes on 4-qubit and 6-qubit problems.
We configure the depth of
the ansatzes so that QAOA and Two-local have 8 and 6 parameters
for $n = 4$ and $n = 6$.}
\label{tab:compare_recon_error_qaoa_twolocal}
\vspace{-0.1in}
\end{table}

\subsubsection{Metrics}


To evaluate the quality of reconstructed landscapes, 
we use the normalized root-mean-square error (NRMSE), a scale-invariant error metric to analyze a diverse set of landscapes. For $x$, the original landscape flattens to a 1-D array, and $y$ the reconstructed landscape flattens to a 1-D array with $T$ elements in total, NRMSE is calculated as

\begin{equation}
\mathrm{NRMSE} = 
{\sqrt {\frac {\sum _{t=1}^{T}({x}_{t}-y_{t})^{2}}{T}}}
\bigg/ (Q_3 - Q_1),
\end{equation}
where $Q_1$ and $Q_3$ are the first and third quartiles of true landscape $x$.

\begin{figure*}[t]
\centering
\begin{subfigure}[t]{0.5\textwidth}
\centering
\includegraphics[width=0.94\textwidth]{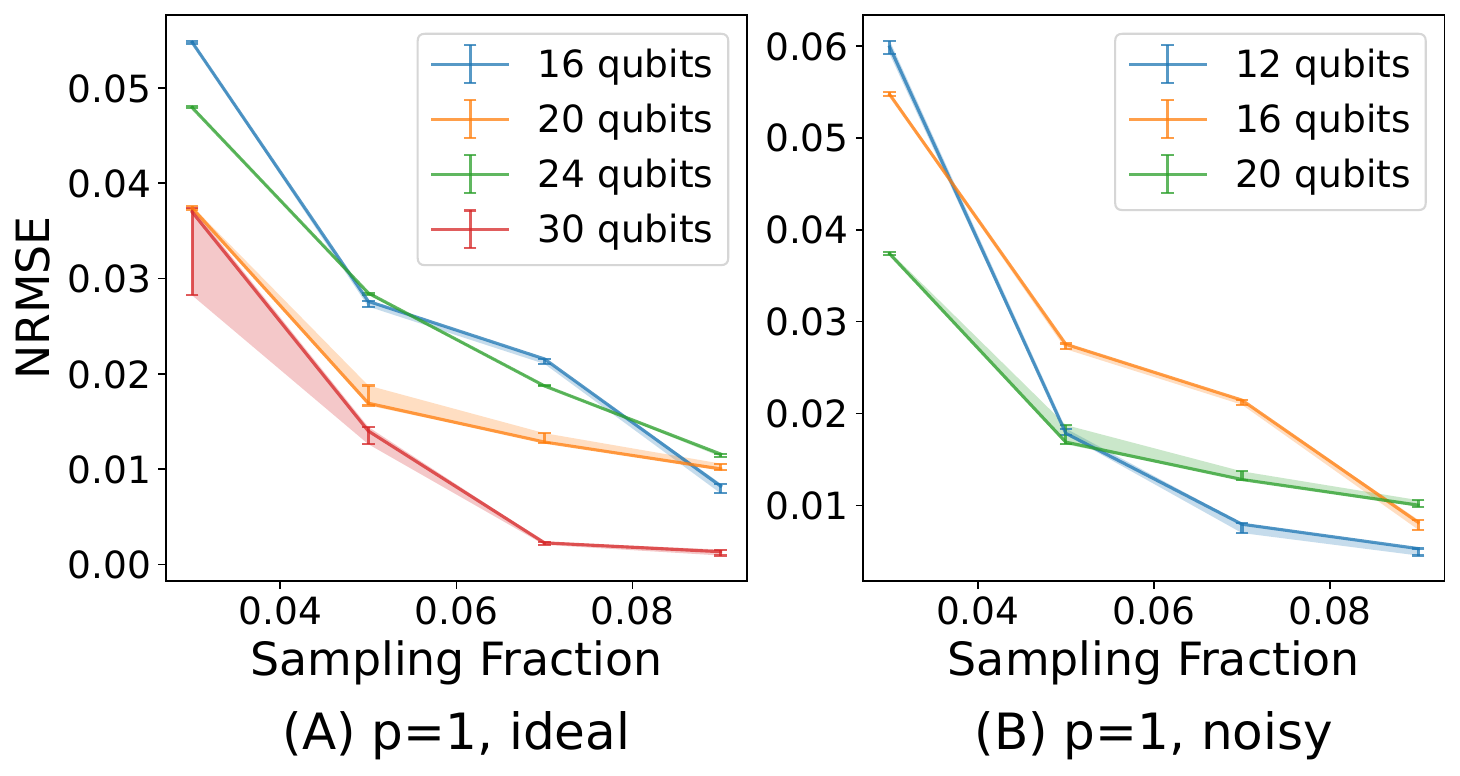}
\end{subfigure}%
~ 
\begin{subfigure}[t]{0.5\textwidth}
\centering
\includegraphics[width=\textwidth]{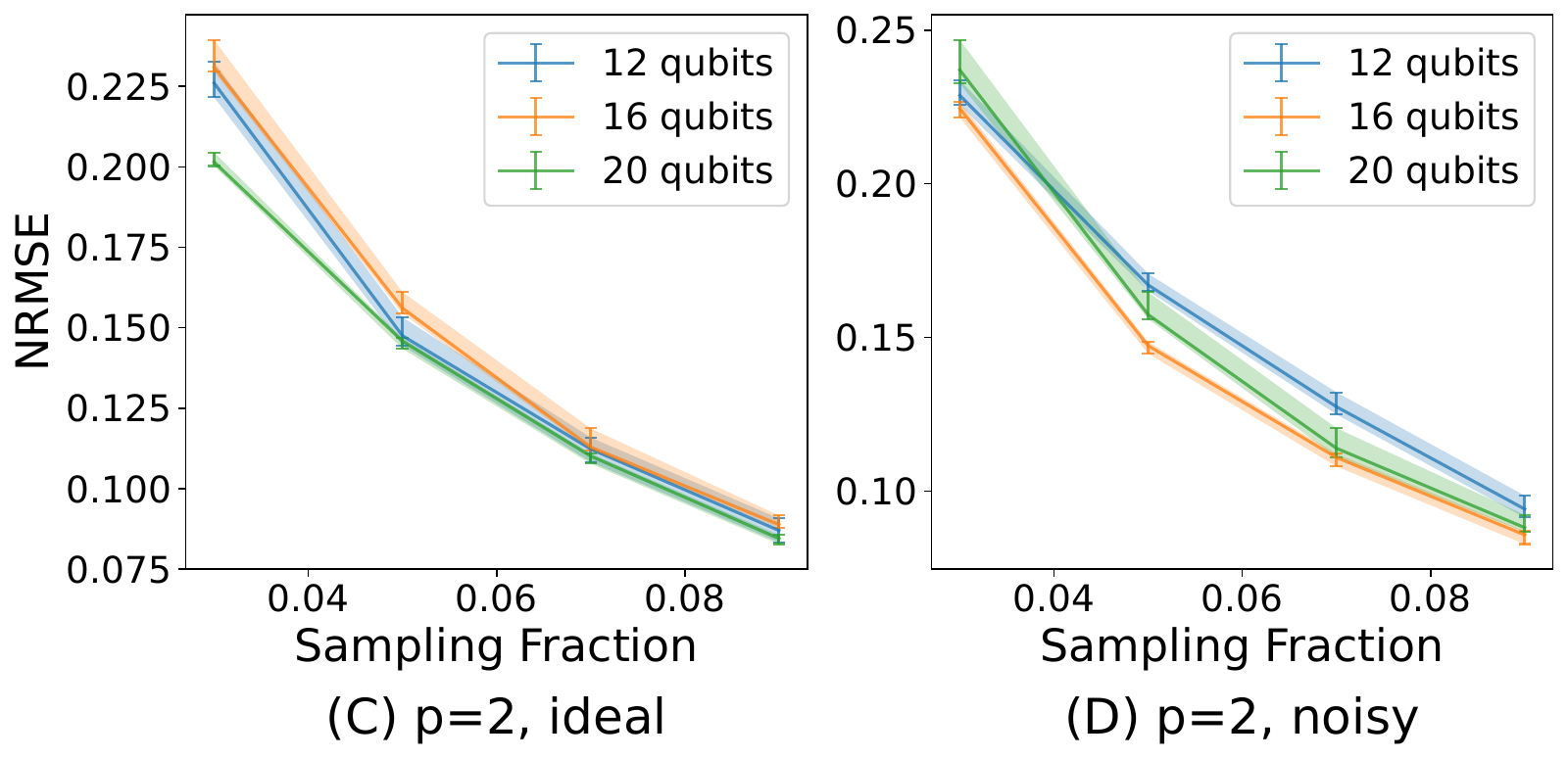}
\end{subfigure}
\caption{Median reconstruction errors of 16 MaxCut problem landscapes for {\sc (a) qaoa}  with $p=1$ and ideal circuit execution  {\sc (b) qaoa} with $p=1$ and depolarizing noise, {\sc 1q} gate error of 0.003 and {\sc 2q} gate error of 0.007.  {\sc (c) qaoa} with $p=2$ and ideal circuit execution  {\sc (d) qaoa} with $p=2$ and depolarizing noise, {\sc 1q} gate error of 0.003 and {\sc 2q} gate error of 0.007.}
\label{fig:recon_error_AB}
\end{figure*}


\begin{figure*}[h]
\centering
\includegraphics[width=1.00\linewidth]{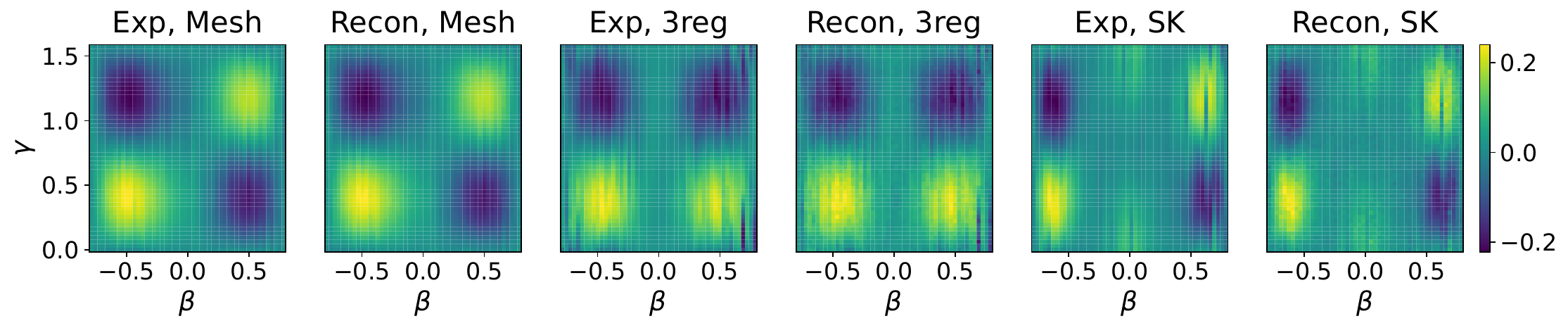}
\caption{Original experimental~\cite{harrigan_quantum_2021} and reconstructed QAOA landscapes for MaxCut problem with mesh graph and 3-regular graph, and SK Model problem. The reconstructed landscapes are generated with a sampling fraction of
41\%.}
\vspace{-0.1in}
\label{fig:google_data}
\end{figure*}

\subsubsection{Evaluating OSCAR with diverse Anzatzes and problems}

While our primary focus is low-depth QAOA and the well-studied 3-regular graph MaxCut problem, our method is applicable to landscapes of a variety of ansatzes and problems. In addition to large-scale QAOA-MaxCut experiments shown in the rest sections, we perform small-scale experiments computing the landscapes of the SK model problem, the hydrogen molecule, and the lithium hydride molecule. We test the QAOA ansatz, the hardware-efficient Two-local ansatz, and the chemistry-inspired UCCSD ansatz, as shown in Table~\ref{tab:compare_recon_error_qaoa_twolocal} and Table~\ref{tab:recon_error_chem}. We configure the depth of the ansatz so that QAOA and Two-local have 8 and 6 parameters for $n=4$ and $n=6$, respectively. Hydrogen and lithium hydride with UCCSD have 3 and 8 parameters respectively. For 8-parameter instances, we sample 7 points along each dimension of the grid; for 3-and-6-parameter instances, we sample 14 points. Due to the exponentially large overhead, we evaluate the reconstruction accuracy by randomly selecting two varying parameters, fixing the rest to random values, and repeating 100 times.

To demonstrate these landscapes are periodic enough to be accurately and efficiently reconstructed, we employ Discrete Cosine Transformation (DCT) to show they are sparse in the frequency domain. Table~\ref{tab:dct_coeff} shows the fractions of DCT coefficients needed to preserve 99\% of the signal energy. We observe that only a few DCT terms contribute to most of the energy costs, showing a highly sparse structure in the frequency domain. Our insight is broadly applicable to VQA landscapes. 


\begin{table}[b]
\renewcommand{\arraystretch}{1.8}
\centering
\setlength{\tabcolsep}{0.05cm}
\renewcommand{\arraystretch}{1.2}
\resizebox{0.82\columnwidth}{!}{%
\begin{tabular}{cccc c c}
\toprule
Molecule & Ansatz   & \#Qubits & \#Parameters & \#Samples & NRMSE \\
\midrule
H$_2$    & Two-local & 2 & 4 & 14        & 0.171 \\
LiH      & Two-local & 4 & 8 & 7         & 0.678 \\
H$_2$    & UCCSD    & 2 & 3 & 14        & 0.345 \\
H$_2$    & UCCSD    & 2 & 3 & 50        & 0.005 \\
LiH      & UCCSD    & 4 & 8 & 7         & 0.856 \\
\bottomrule
\end{tabular}%
}
\caption{Reconstruction errors of hydrogen and lithium hydride molecules. ``\#Samples" indicates the number of points equidistantly sampled within the range of every parameter.}
\label{tab:recon_error_chem}
\end{table}

\begin{table}[b]
\centering
\resizebox{\columnwidth}{!}{%
\begin{tabular}{cccc}
\toprule
      Problem           & QAOA Ansatz    & Two-local Ansatz & UCCSD Ansatz\\
\midrule
 3-reg Maxcut (n=4)     & 0.0420\%       & 0.0000867\%     & --\\ 
 3-reg Maxcut (n=6)     & 0.00768\%      & 0.0000133\%     & --\\ 
 SK Problem (n=4)       & 0.0420\%       & 0.000416\%      &--\\ 
 SK Problem (n=6)       & 0.00912\%      & 0.0000398\%     & --\\ 
 H$_2$ (n=2)            & --      & 0.00260\%       & 0.0729\%     \\
 LiH (n=4)              & --       & 0.000104\%      & 0.0000173\%     \\
\bottomrule
\end{tabular}%
}
\caption{The fractions of DCT coefficients needed to preserve 99\% of the signal energy, verifying that VQA landscapes are generally sparse in the frequency domain and thus can be accurately and efficiently reconstructed.}
\label{tab:dct_coeff}
\end{table}

\subsubsection{Evaluations on QAOA}

Figure~\ref{fig:recon_error_AB} shows the accuracy of landscape reconstruction versus sampling fractions for different QAOA problem. For ideal and noisy cases, we reconstruct landscapes of 16 different MaxCut problems and draw 25\%, 50\%, and 75\% quartiles for given numbers of qubits under varying sampling fractions. As the sampling fraction increases, the reconstruction error of both ideal and noisy decreases steadily and keeps having small variances across instances.   Figure~\ref{fig:recon_error_AB} (C) and (D) show the error of reconstructing $p=2$ ideal and noisy landscapes. We evaluate the reconstruction accuracy of the $p=2$ case by reshaping 4-D landscapes into 2-D and then applying OSCAR. For example, we reshape the true landscape with dimensions (12,12,15,15) into a (12$\times$12, 15$\times$15) landscape.
We observe that despite the noisy operations or higher dimensionality, OSCAR can perform accurate landscape reconstruction, capturing the impact of noise on the landscape. However, with increased dimensionality, the reconstruction accuracy drops because of reshaping, which introduces artificial repeating patterns. Moreover, simulating $p=2$ noisy landscapes is slow\footnote{To generate the entire optimization landscape, we need to simulate 32k 20-qubit depth-2 MaxCut QAOA circuits. On the Qiskit state vector simulator using a single Nvidia A100 GPU, this requires 45 to 55 hours of runtime.}.

\subsection{Evaluating OSCAR with Google's Dataset}

\begin{figure}[h]
\centering
\includegraphics[width=0.65\linewidth]{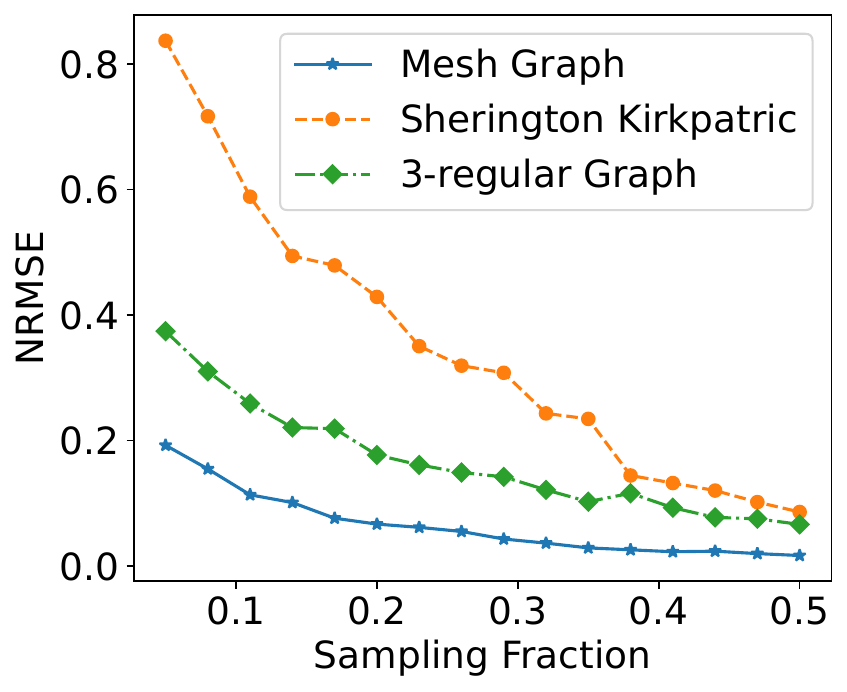}
\caption{Reconstruction error on Google Sycamore}
\vspace{-0.2in}
\label{fig:google_recon_error}
\end{figure}

To evaluate the reconstruction accuracy of OSCAR on real hardware with complex noise sources, we utilize the QAOA dataset from Google~\cite{harrigan_quantum_2021}. This includes the optimization landscapes generated on Google's 53-qubit Sycamore chip for the MaxCut on the 3-regular graph, MaxCut on the mesh graph, and the Sherington Kirkpatric (SK) Model problem~\cite{sherrington_solvable_1975}. All landscapes in the Google dataset have $50 \times 50$  data points. Figure~\ref{fig:google_data} summarizes reconstruction errors with increasing sampling fraction. Note that Google's original landscape has 2500 points, which is relatively sparse compared to dense grids with 5000 to 32,000 points that we have used previously. We believe due to the sparser original landscape, the reconstruction accuracy is slightly lower. However, the error measured by the NRMSE is conservative as shown in Figure~\ref{fig:google_recon_error}. For the NRMSE of about 0.2, the reconstructed landscapes are perceptually identical to the original landscapes. Figure~\ref{fig:google_recon_error} highlights successful reconstruction for the SK Model landscape, despite the highly noisy original landscape.  

\begin{tcolorbox}
\textbf{Using a small number of samples, OSCAR can accurately reconstruct the landscape by providing 2x to 20x speedups for complete landscape generation.}   
\end{tcolorbox}
\vspace{-0.1in}






\begin{figure*}[t]
\centering
\begin{subfigure}[b]{\textwidth}
\centering
\includegraphics[width=\textwidth]{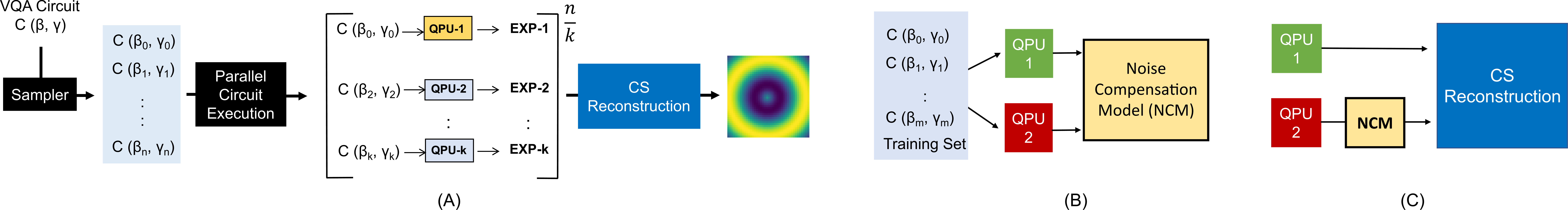}
\label{fig:diagram_parallel_1}
\end{subfigure}\hspace{4mm}
~ 
~ 
\vspace{-0.2in}
\caption{(A) OSCAR in Parallel Mode (B) Training Noise Compensation Model (NCM)  to map expected cost values obtained on QPU-2 to QPU-1 (reference machine) (C) Leveraging NCM to transform QPU-2 values to enable noise-preserving reconstruction.}
\label{fig:diagram_parallel}
\vspace{-0.1in}
\end{figure*}

%% file: sections/5OSCAR_D.tex

\section{Parallel Landscape Reconstruction}

Generating a full landscape requires thousands of shots for a large number of points in the circuit parameter space. With OSCAR, we reduce the number of experiments significantly by using compressive sensing.
Furthermore, we can accelerate the landscape generation even further by exploiting the parallelism offered by OSCAR. When debugging, OSCAR decouples the optimizer to eliminate the sequential execution model. In this section, we will discuss parallel landscape reconstruction, a design that takes advantage of the parallel circuit execution as the samples required can be generated independently. 

\subsection{Parallel Reconstruction with OSCAR}

Figure~\ref{fig:diagram_parallel} shows OSCAR can generate cost function samples in parallel by executing quantum circuits on \textit{k} QPUs as samples picked on the cost function landscape are entirely independent.  
This constant factor speedup of  \textit{k}  can significantly help the debugging process. Due to scarce quantum hardware and costly quantum simulations,  users may have to wait for long durations to obtain the results when debugging quantum programs. For example, on publicly available QPUs, the queuing delay for running hundreds of circuits can span from a few hours to days~\cite{gokul,das2019case}. Even when using quantum simulations, runtimes can be extremely long. Using OSCAR, users can leverage multiple quantum computers (or simulators) at once. However, when we use samples generated on multiple QPUs, the reconstructed landscape can be a `mixture' of all the landscapes that we are sampling from, which might be acceptable for specific debugging use cases. But for a use case where the user wants to study the impact of the hardware error on the VQA workflow, preserving the effect of noise on the reconstructed landscape is necessary. 
Note that prior works have used circuit ensembles to boost fidelity and performance ~\cite{tannu_ensemble_2019,tannu2019mitigating,stein2022eqc,resch2021accelerating}. However, using them as is may not be apt for debugging VQAs. 

\begin{figure}[b]
\centering
\vspace{-0.2in}
\includegraphics[width=0.95\linewidth]{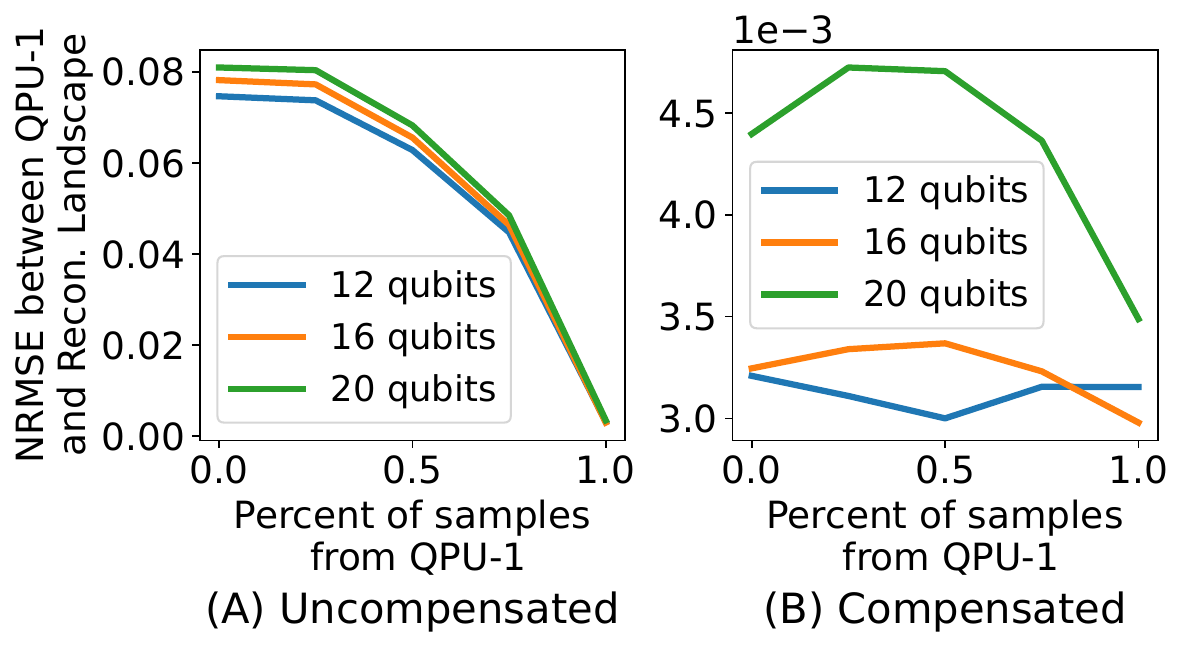}
\caption{Errors between reconstructed and target (QPU-1's) landscapes using samples from two QPUs, without (A) and with NCM (B). }
\label{fig:parallel_LS_recon_normalized}
\end{figure}

To enable noise-preserving reconstruction in a multi-QPU setting, we propose a {\em Noise Compensation Model (NCM)} based on linear regression. First, we train NCM on a small fraction of circuit parameter samples executed on QPU-1 and QPU-2 as shown in Figure~\ref{fig:parallel_LS_recon_normalized}(B). Then, using NCM, we can transform the expected values on QPU-2 to match the noise configuration of reference machine QPU-1 as shown in Figure~\ref{fig:parallel_LS_recon_normalized}(C). To illustrate the effectiveness of our method, we run three QAOA problems on two QPUs with different noise configurations. The 1Q and 2Q gate error rates on QPU-1 are 0.1\% and 0.5\%, respectively, whereas, for  QPU-2, the error rates are 0.3\% and 0.7\%. Our goal is to match the landscape obtained on QPU-1 by using a mixture of QPU-1 and QPU-2 samples. Figure~\ref{fig:parallel_LS_recon_normalized} shows the NRMSE between the reconstructed landscape that uses samples from QPU-1 and QPU-2, while the true landscape obtained on QPU-1 is used as the reference. When OSCAR is executed in uncompensated mode, the reconstruction error is high, indicating dissimilarity between the reconstructed landscape and the target landscape. However, by using NCM, OSCAR can significantly reduce the reconstruction error and closely match the target landscape as shown in Figure~\ref{fig:parallel_LS_recon_normalized}(B). For training NCM, we use 1\% training samples from the landscape, while for reconstruction, a total of 10\% samples with a varying fraction from QPU-1 and QPU-2 are used.

Moreover, we test the performance of NCM on real quantum hardware by introducing samples generated on IBM Lagos and IBM Perth.
Table~\ref{tab:parallel_recon} shows the reconstruction errors of different combinations of IBM devices and simulations as sources, with and without NCM. We observe that NCM makes the reconstruction landscape closer to the target in all cases. Furthermore, there is little difference between using ideal and noisy simulation as QPU2. This result indicates that with a small fraction of samples from real quantum devices and the rest from ideal simulation, we can accurately reconstruct landscapes generated on real quantum devices.

\begin{table}[h]
\renewcommand{\arraystretch}{1.8}
\centering
\setlength{\tabcolsep}{0.03cm}
\renewcommand{\arraystretch}{1.1}
\resizebox{1.05\linewidth}{!}{%
\begin{tabular}{ccccccccc}
\toprule
    \multirow{2}{*}{QPU1} &
    \multirow{2}{*}{QPU2} &
      \multicolumn{2}{c}{20\%-80\%} &
      \multicolumn{2}{c}{50\%-50\%} &
      \multicolumn{2}{c}{80\%-20\%} &
      \multicolumn{1}{c}{100\%-0\%} \\
      & & {\sc oscar} & {\sc+ncm} & {\sc oscar} & {\sc +ncm} & {\sc oscar} & {\sc +ncm} & {\sc oscar} \\
\midrule
{\sc n}oisy {\sc s}im-{\sc i} & {\sc n}oisy {\sc s}im-{\sc ii}  & 0.076             & 0.003                 & 0.061             & 0.002                 & 0.039             & 0.002                 & 0.001           \\
{\sc n}oisy {\sc s}im-{\sc ii} & {\sc n}oisy {\sc s}im-{\sc i}  & 0.075             & 0.002                 & 0.059             & 0.002                 & 0.037             & 0.002                 & 0.001           \\
{\sc ibm} {\sc p}erth         & {\sc i}deal {\sc s}im         & 1.362             & 0.299                 & 0.97              & 0.265                 & 0.597             & 0.223                 & 0.184           \\
{\sc ibm} {\sc p}erth         & {\sc n}oisy {\sc s}im         & 0.767             & 0.272                 & 0.564             & 0.247                 & 0.379             & 0.213                 & 0.184           \\
{\sc ibm} {\sc p}erth         & {\sc ibm} {\sc l}agos         & 0.5               & 0.424                 & 0.419             & 0.36                  & 0.284             & 0.262                 & 0.184           \\
{\sc ibm} {\sc l}agos & {\sc ibm} {\sc p}erth & 0.403 & 0.337 & 0.341 & 0.286 & 0.266 & 0.247 & 0.222  \\
{\sc i}deal {\sc s}im & {\sc ibm} {\sc p}erth & 0.478 & 0.226 & 0.363 & 0.18 & 0.215 & 0.109 & 0.042 \\
\bottomrule
\end{tabular}%
}
\caption{
Errors between reconstructed and QPU1 landscapes using different combinations of IBM devices and simulations as sources, with and without the Noise Compensation Model (NCM). ``+NCM" indicates the samples from QPU2 are transformed to match the noise model of QPU1 (target landscape). ``(20\%-80\%)" means 20\% of the samples come from QPU1 and 80\% come from QPU2.
}
\label{tab:parallel_recon}
\vspace{-0.3in}
\end{table}


\subsection{Relaxing Amdahl's Law with Eager Reconstruction} Generating an entire landscape is embarrassingly parallel. However, as per Amdahl's law, the serial portion fundamentally limits the parallel speedup. This applies to circuits running on parallel QPUs. Recent data suggest that a significant fraction of the wait times on quantum computers are due to queuing delays~\cite{gokul}. Furthermore, during our evaluations, we see 10x to 30x higher tail latency than median circuit evaluation latencies. With the increasing complexity, we expect higher queuing delays and large tail latencies, making debugging challenging on future quantum computers. However, we can augment OSCAR to sidestep Amdahl's law. We propose \textit{eager reconstruction} for OSCAR, where we leverage the tradeoff between reconstruction accuracy and sampling fraction. Based on our evaluations, we observe that a small reduction in sampling fraction does not reduce reconstruction accuracy significantly. As a result, during the circuit sampling process, we can set a soft timeout and begin the eager reconstruction for the available samples that fall under this timeout. In short, we can omit the samples that incur tail latency to significantly reduce the time to reconstruct without sacrificing the reconstruction accuracy.






%% file: sections/6ucase-1.tex
\section{Use Case: Benchmarking and Tuning Noise Mitigation}

Noise mitigation methods are shown to be effective in reducing the errors introduced by the current noisy quantum hardware~\cite{kandala2019error,russo2022testing}. However, deciding which mitigation method to use is not straightforward, and comparing the performance of different mitigation methods and settings can be highly resource-consuming. We show that reconstructed landscapes with OSCAR preserve properties of the original landscapes while being easily computable, thus helping us benchmark and configure noise mitigation techniques in an intuitive manner at a substantially reduced cost.

Inspired by the recent demonstration of Zero Noise Extrapolation (ZNE)~\cite{li_efficient_2017} on multiple quantum hardware platforms~\cite{russo2022testing}, we use ZNE as an example to demonstrate how OSCAR can help visualize, benchmark, and configure noise mitigation methods. Given a circuit, ZNE first generates a few equivalent circuits by replacing a gate $U$ with equivalent substitutions, such as $UU^{-1}U$, based on the predefined noise scaling factors. The new circuits are thus functionally identical to the original but with additional noise. Then, by running these circuits with varying degrees of noise, we calculate the expected cost function values at different noise levels and then estimate expected noise with zero noise by using user-configured extrapolation models. ZNE is costly to apply since it requires running a significant number of additional circuits. Furthermore, it is even more expensive to configure, as there are various extrapolation models and many scaling factor choices.

\begin{figure}[t]
\centering
\includegraphics[width=0.75\linewidth]{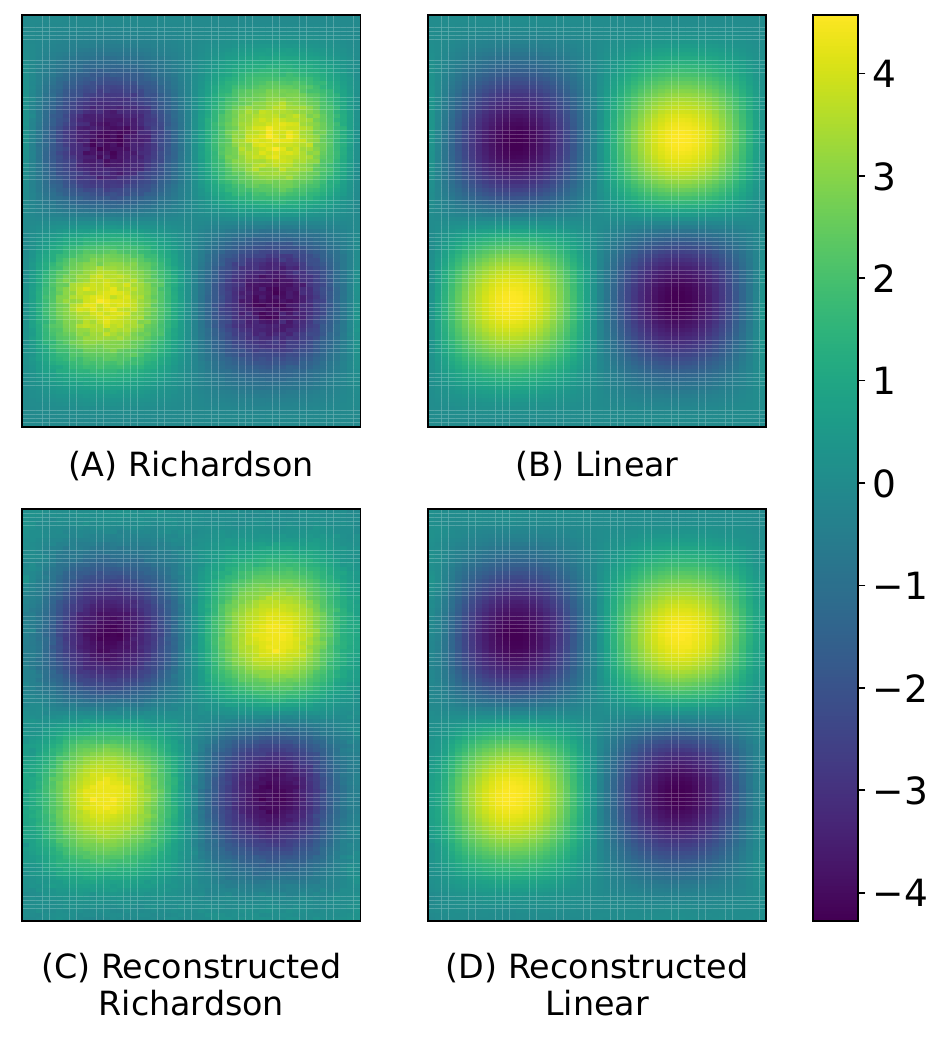}
\caption{Richardson extrapolation and linear extrapolation
mitigation methods on a depth-1, 16-qubit landscape with
depolarizing noise ({\sc 1q} error of 0.001 and {\sc 2q} error of 0.02.}
\vspace{-0.25in}
\label{fig:comp_Richardson_and_Linear}
\end{figure}

Noise mitigation configuration can strongly influence the overall efficacy of mitigation techniques. For example, we use the Richardson extrapolation~\cite{li_efficient_2017,temme_error_2017} with \{1,2,3\} scaling and the linear extrapolation with \{1,3\} scaling for solving MaxCut problem using QAOA, and  Figure~\ref{fig:comp_Richardson_and_Linear} shows a landscape comparison of Richardson and linear extrapolation models. We can observe that Richardson extrapolation has a salt-like noise, whereas linear extrapolation leads to a smoother landscape closer to the ideal landscape almost without any noise. Richardson extrapolation adds a jaggedness to the landscape, which makes gradient-based optimization difficult.

As shown in Figure~\ref{fig:comp_Richardson_and_Linear}, the reconstructed landscapes preserve this difference. Thus OSCAR can enable users to study the impact of noise mitigation on the optimization landscape. Furthermore, we can quantitavely measure and compare the performance of different configurations. For example, we want to evaluate how mitigated features such as roughness and flatness change with different noise mitigation configurations using the following metrics:

\vspace{0.01in}

{\noindent (a) second-order derivative}
\begin{equation}
\mathrm{D^2}(x)=\sum_i\left[x_i-2 x_{i-1}+x_{i-2}\right]^2 / 4,
\end{equation}
(b) variance of gradients on the landscape
\begin{equation}\label{eq:VoG}
\mathrm{VoG}(x) = \mathrm{Var}\left[x_i - x_{i-1}\right]
\end{equation}
and (c) variance of the landscape
\begin{equation}
\sigma^2(x) = \mathrm{Var}\left[x_i\right].
\end{equation}

A second-order derivative on a landscape is commonly used to evaluate the roughness~\cite{green_nonparametric_1993}. The variance of gradients is typically used to measure barren plateaus or flatness of the landscape~\cite{mcclean_barren_2018}. Eq.~\ref{eq:VoG} expresses gradients with the difference of elements. Note that all the equations above are for one-dimensional $x$. We compute average metrics on all dimensions.

\begin{figure*}[t]
\centering
\includegraphics[width=0.9\linewidth]{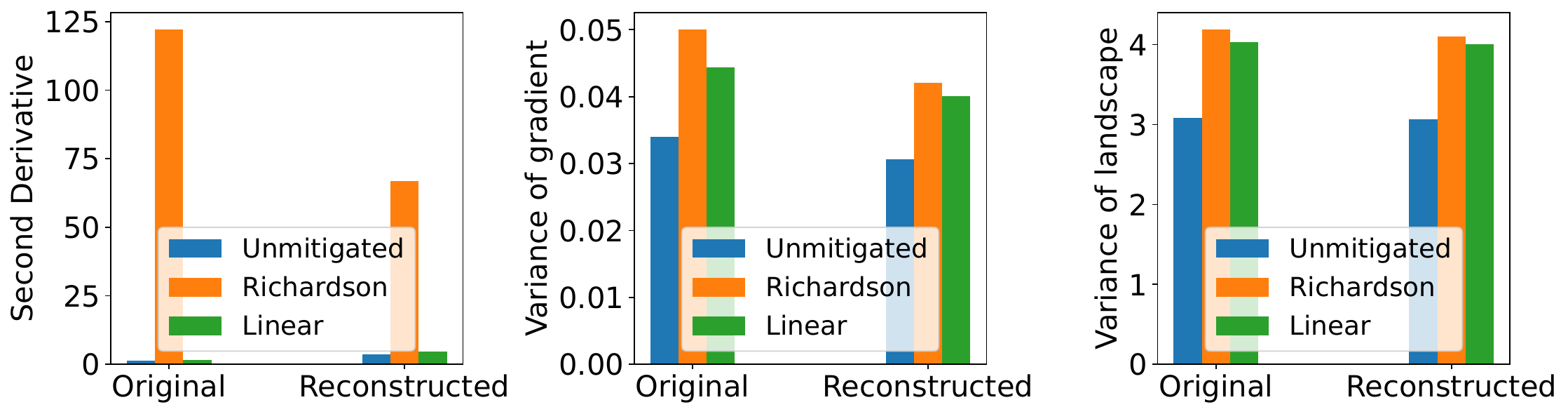}
\caption{Reconstructed landscapes using OSCAR preserve key features of the original landscapes.}
\label{fig:miti_preserve}
\vspace{-0.05in}
\end{figure*}

We compare the three metrics of the original and the reconstructed landscape calculated with unmitigated results and Richardson-and-linear-extrapolated ZNE results in Figure~\ref{fig:miti_preserve}. 
The reconstructed landscapes preserve the features of the original landscapes, enabling users to infer the impact of noise mitigation on the overall landscape. With OSCAR, we can evaluate landscape properties with substantially less costly reconstructed landscapes, which are derived with only a small fraction of points from the original landscapes.

As shown in Figure~\ref{fig:miti_preserve}, ZNE with linear or Richardson extrapolation has comparable average flatness metrics as VoG and Variance values on the original landscape are comparable. A similar conclusion can be drawn by using OSCAR, where reconstructed landscapes show a similar trend. Unlike flatness, the roughness of Richardson's is significantly different as the second derivative is extremely high on the reconstructed landscape, which agrees with the ground truth of the original landscape, shown in Figure~\ref{fig:comp_Richardson_and_Linear}.

%% file: sections/6ucase-4.tex
\section{Use Case: Configuring and Debugging Optimizer}

 VQAs, being heuristic algorithms and containing several configurable components, require extensive tuning to work as expected on a given problem. The classical optimizer in the VQA workflow is arguably the most tricky component to set up.
Assume we are trying to configure the optimizer, and test the QAOA workflow for a large problem that is expensive to run on the QPU and difficult to simulate using a classical computer. However, when we begin the QAOA training process, the optimizer is not able to make significant progress even after running the training loop a significant number of times.


In the current quantum software development paradigm, even to test the configurations of an optimizer it is imperative to run the circuit a large number of times with different configurations. 
The trial and error approach of debugging may not always work as the desired outcome of the VQC training process may be hindered by multiple reasons such as bad optimizer configurations, hardware errors, software bugs, etc. 
Moreover, when you run the optimizer it only gives you the local information which is the path traversed by the optimizer in a small parameter range. Ideally, we would like to have a bird’s-eye view and see the full landscape and how optimizers make progress to get more insights into the problem. This is essential for near-term variational quantum algorithms as writing and debugging quantum programs are not very intuitive. 
Unlike classical programs which have meaningful intermediate states that programmers can use to debug their implementation of the algorithm, debugging VQAs by running quantum circuits with small sets of arbitrary parameters are hard to reason with. Furthermore, VQCs are known to have barren plateaus on the landscape, where cost becomes insensitive to the changes in the circuit parameters. Similar to the vanishing gradient problem~\cite{pascanu_difficulty_2012} in machine learning, searching for the optimal parameters using the gradients becomes challenging. Barren plateaus can cause the optimizer to get stuck in a local region of the parameter space. In addition to barren plateaus, it may be possible to encounter saddle points and local minima, making optimization non-trivial.

\begin{figure}[b]
\centering
\includegraphics[width=0.85\linewidth]{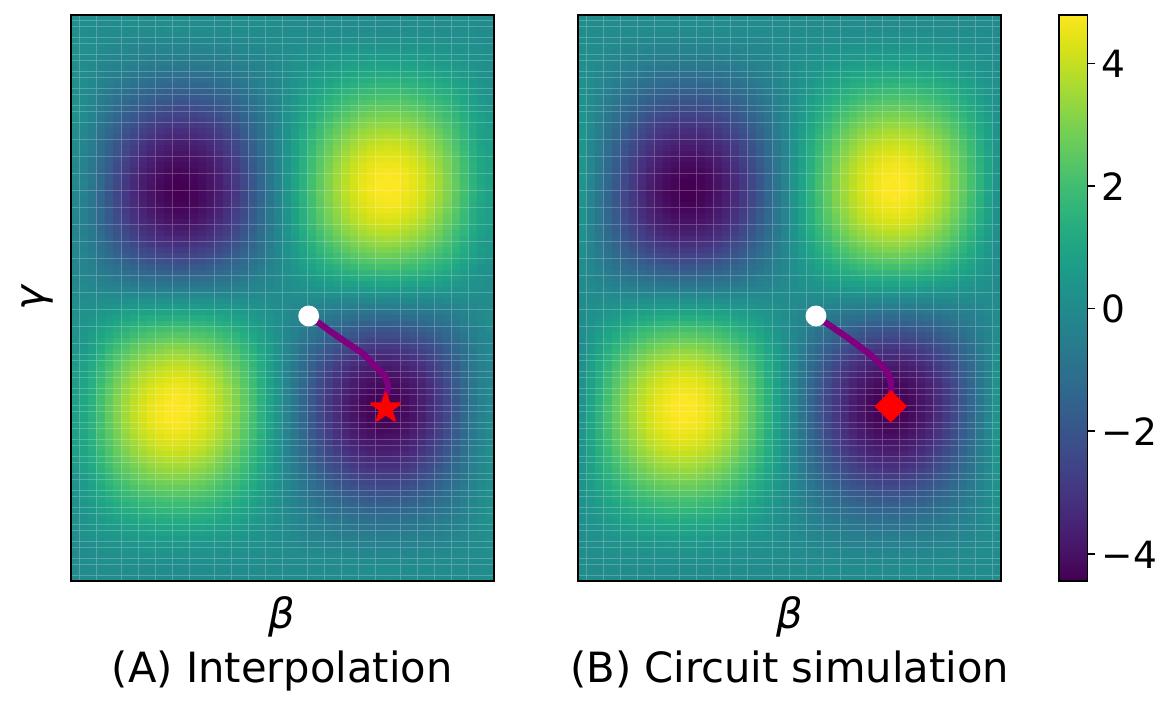}
\caption{(A) Optimization on the reconstructed landscape by ADAM. (B) Optimization by circuit simulation. The landscape is from a 16-qubit MaxCut instance.}
\label{fig:compare_intp_and_circ}
\end{figure}
\begin{figure}[t]
\centering
\includegraphics[width=0.7\linewidth]{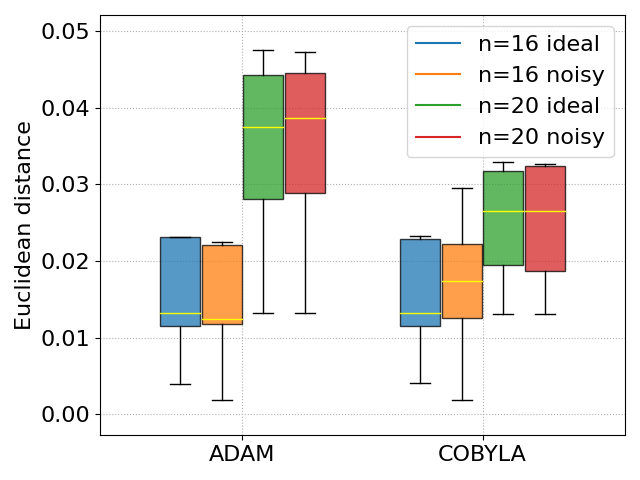}
\caption{Comparison of the Euclidean distances between the ending points of optimizing on 
the reconstructed landscape and with circuit executions.}
\vspace{-0.1in}
\label{fig:optimization_box}
\end{figure}

Running the optimizer on the reconstructed landscape can serve as a good way to pre-check the optimizer performance before actually executing circuits on a real quantum device. To allow continuous-space optimization on the discrete grid, we use rectangular bivariate spline interpolation~\cite{nurnberger_bivariate_1995} to fill in the gaps. Figure~\ref{fig:compare_intp_and_circ} shows an example comparison between optimizing on the interpolated reconstructed landscape and with circuit executions, where the optimizer paths are identical. To evaluate the efficacy of our method, we calculate the Euclidean distance between the ending points of the two optimization paths. We use the gradient-based optimizer ADAM and the gradient-free optimizer COBYLA with default settings from \verb|Qiskit| and random initial points.
We run 8 instances each for ideal and noisy simulations of 16-qubit and 20-qubit problems. Figure~\ref{fig:optimization_box} shows that interpolated reconstructed landscapes and circuit executions result in very close endpoints.

After obtaining reconstructed landscapes, our method is significantly faster and more resource-efficient compared to the normal optimization, since interpolation returns an optimizer query almost instantly. For some optimizers, the number of QPU queries required by reconstruction can even be fewer than the number needed for actual optimizations. In addition, our method allows reruns of optimizations with little overhead after reconstructing the landscape. Figure~\ref{fig:compare_ADAM_and_COBYLA} shows an example of choosing and configuring optimizers based on their performance on the reconstructed landscape.

%% file: sections/6ucase-3.tex
\section{Use Case: Initializing Optimizer using OSCAR}

Selecting accurate initialization for VQA circuit can have a considerable impact on the optimization result. For many problems, users have to resort to randomly choosing an initial point for the optimizer,  leading to large number of cost function queries and worse results than using a properly chosen initial point. Past studies have suggested warm-starting with classical algorithms as well as using parameters obtained from running simpler instances~\cite{egger2021warm}. 

We propose an alternative approach that takes advantage of the reconstructed landscape. Since parameters obtained by optimizing on the interpolated reconstructed landscape are very close to the optimal ones, as shown by Figure \ref{fig:optimization_box}, they are suitable choices as initial points for the regular optimization workflow. Using OSCAR, we generate a reconstructed landscape, interpolate it, and run an optimizer to obtain the minima on the reconstructed landscape. We use the minima that OSCAR converges to as the initial point to run optimization using regular workflow.

We run depth-1 QAOA on 14 different instances of the 16-qubit MaxCut problem with ideal and noisy simulators, starting from randomized initial points and points given by optimizing the reconstructed landscape, respectively. The resulting cost function values are within the termination tolerance of the optimizers, but OSCAR-generated initial points require much fewer queries during the optimization process. Even with the additional queries needed for reconstructing the landscapes, OSCAR can still be significantly faster for optimizers such as ADAM, as shown in the first two rows of Table~\ref{tab:qpu_queries}. 
For optimizers that require fewer queries in nature, such as COBYLA, OSCAR is slower in paying the overhead of landscape reconstruction queries, as shown in the last two rows of Table~\ref{tab:qpu_queries}.
However, we stress that these queries can be computed completely in parallel. In that case, OSCAR reduces the sequential QPU computation time but adds an overhead for performing compressed sensing. The aforementioned concurrent study~\cite{fontana_efficient_2022} proposes a similar approach and tests with gradient descent in the ideal setting, where they select the minima of the reconstructed landscape as initial points. While we agree with the claim that such reconstructed landscape-based initialization can greatly enhance gradient-based optimizers, we also note that it may not always be cost-effective to use our method for selecting initial points, especially for gradient-free optimizers such as COBYLA. 

\begin{figure}[t]
\centering
\includegraphics[width=0.8\linewidth]{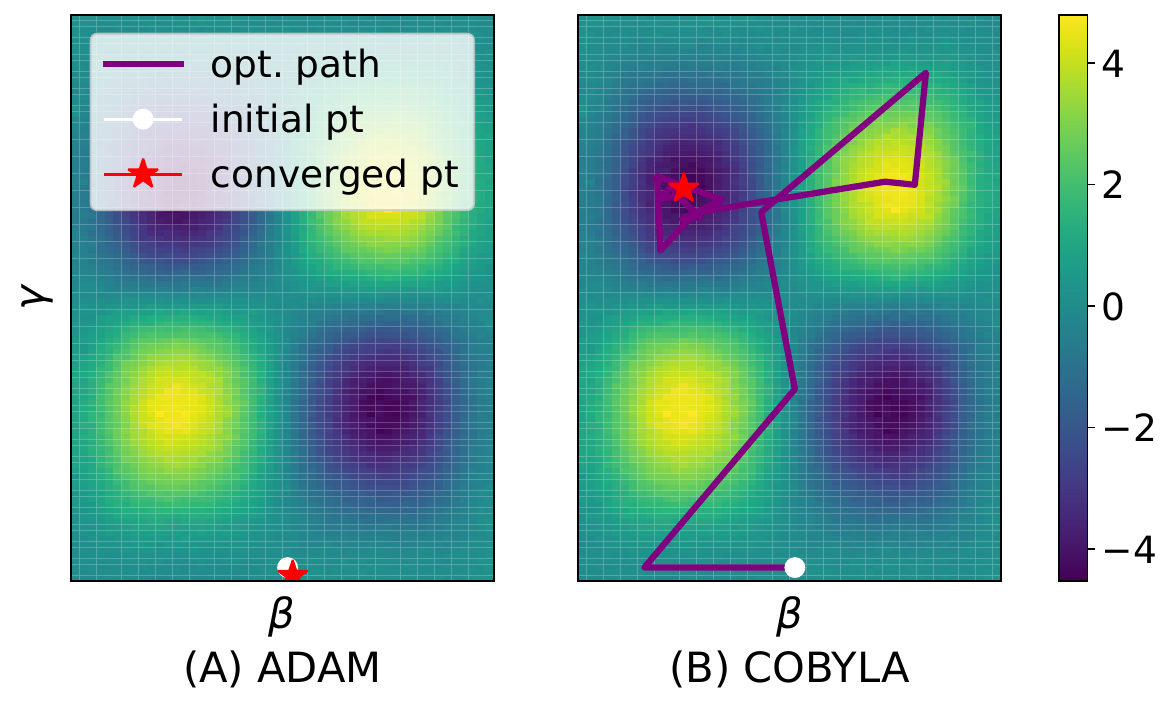}
\caption{ An example of choosing the optimizer based on the interpolated reconstructed optimization. On this Richardson extrapolated landscape, a gradient-free optimizer (COBYLA) performs better than a gradient-based one (ADAM).}
\label{fig:compare_ADAM_and_COBYLA}
\vspace{-0.20in}
\end{figure}


\begin{table}[t]
\centering

\resizebox{\columnwidth}{!}{%
\begin{tabular}{cccc}
\toprule
              & random, opt. & OSCAR, opt. & OSCAR, opt.+recon. \\
\midrule
ADAM, ideal   & 3127        & 370        & 620                 \\
ADAM, noisy   & 3123        & 661        & 911                 \\
COBYLA, ideal & 38          & 32         & 282                 \\
COBYLA, noisy & 40          & 32         & 282                 \\
\bottomrule
\end{tabular}%
}
\vspace{0.1in}
\caption{Number of QPU queries to reach convergence by ADAM and COBYLA running depth-1 QAOA on 14 instances of the 16-variable MaxCut problem, using OSCAR and random initialization, respectively.
``opt." is the number of queries during optimization,
and ``recon" is the number of queries during landscape reconstruction.}



\label{tab:qpu_queries}
\vspace{-0.3in}
\end{table}

%% file: sections/8RelatedWorks.tex
\newpage
\section{Related Work}\label{sec:related_work}
{\noindent \textbf{Debugging Quantum Programs:}} Debugging quantum software is an open problem, and researchers have just started to study bugs in quantum software~\cite{paltenghi2022bugs,campos2021qbugs,zhao2021bugs4q}. Prior works have focused on assertions to debug quantum programs, which use statistical tests based on classical measurements.  To determine the correctness, prior work uses the chi-square test and contingency table analysis~\cite{huang_statistical_2019}. However, estimating quantum states requires repeat execution of circuits for every assertion point. To reduce the assertion cost, researchers proposed runtime assertion by performing testing and debugging on ancilla qubits~\cite{liu_quantum_2020}, introducing dynamic and indirect verifications, and \verb|Proq|,  a projection-based runtime assertion \cite{li_projection-based_2020}, leveraging those projective measurements that will not change the tested state if the state meets the assertion. Besides, some quantum programming languages, such as \verb|Q#|~\cite{svore_q_2018} and \verb|Cirq|~\cite{cirq_developers_2022_6599601}, provide unit tests and assertions APIs. 


{\noindent \textbf{Periodic Landscapes:}} 
A growing body of research leverages symmetry in QAOA, VQE, and QML landscapes to enable efficient workflows. For QAOA, \cite{shaydulin_exploiting_2021} found a connection between classical symmetries of the objective function and the symmetries of the terms of the cost Hamiltonian with respect to the QAOA energy. Furthermore, it shows that such symmetry is general and broadly applicable independent of input type. For VQE,  symmetry in the landscape naturally arises due to the structure of molecule Hamiltonians. Multiple recent works leverage symmetry of the many-body systems to enable efficient VQE circuits \cite{lyu_symmetry_2023,ryabinkin_symmetry_2018,seki_symmetry-adapted_2020}. Whereas for recent work focusing on QML, \cite{meyer_exploiting_2022,larocca_group-invariant_2022,nguyen_theory_2022} explored symmetry groups of the learning problem to construct models with outcomes invariant under the symmetry of the learning task. For effective error mitigation, multiple prior works~\cite{kakkar_characterizing_2022,shaydulin_error_2021,mcardle_error-mitigated_2019,bonet-monroig_low-cost_2018,streif_quantum_2021}
leverage symmetry to enable effective noise mitigation.

{\noindent \textbf{Concurrent work on VQA landscape reconstruction:}} 
The concurrent work \cite{fontana_efficient_2022} provides theoretical evidence that low-depth QAOA cost landscapes are sparse in the Fourier basis and, thus, are recoverable by CS. 
Empirically, they conduct limited simulated experiments in the ideal setting.
In contrast, we focus on a broader VQA debugging tuning problem. We provide a more thorough and statistically rigorous empirical validation and propose parallel sampling and noise compensation as extensions of this method. We show multiple concrete use cases where OSCAR can be leveraged to accelerate VQA debugging and tuning. 




%% file: sections/9Conclusion.tex
\section{Conclusion}




Debugging VQAs on near-term quantum platforms is extremely challenging due to limited access to the internal state of the quantum program, non-determinism introduced by the optimizer, and noisy operations on near-term quantum computers. In this paper, we propose OSCAR to significantly reduce the overhead of VQA debugging and tuning. OSCAR leverages compressed sensing to generate a complete landscape that provides the user with a bird's-eye view to assess the correctness of all the components involved in executing VQAs. Using simulators and real quantum hardware, we show that OSCAR can generate a complete optimization landscape, executing a small fraction of randomly selected circuits with high accuracy. In addition, by using OSCAR, we break the serial execution model and enable a parallel execution model that can utilize multiple devices for accurate landscape generation and further improve landscape reconstruction. OSCAR can evaluate the efficacy of noise mitigation techniques in an optimizer-agnostic manner and help users to effectively configure noise mitigation on near-term quantum computers. Furthermore, we demonstrate a use case where OSCAR can help troubleshoot wrong optimization configurations. OSCAR can be used to untangle key components of the VQA workflow including the initialization, optimizer, circuit execution, and noise mitigation, to isolate the bugs and help tune the individual components significantly faster and at a lower cost.


%% file: sections/10Appendix.tex
\section{Principle of Compressed Sensing}\label{ap:cs_math}

Here is a detailed mathematical explanation of CS.

\subsection{General Procedure of Compressed Sensing}


Denote $x\in\Rbb^n $ as a compressible vector,
and it could be represented by a sparse vector $s\in\Rbb^n$
in a new basis $\Psi\in \Rbb^{n \times n}$:
\begin{equation}\label{eq:x_eq_Psi_s}
x = \Psi s.
\end{equation}
The sparsity of $s$ is defined as the number of non-zero terms in $s$.
The basis here could be Fourier basis, wavelet basis or other tailored basis.
Every column of $\Psi$ is a base orthogonal to each others.
Eq. \ref{eq:x_eq_Psi_s} is assuming that the original signal $x$
could be represented by very few coefficients in another basis.

Let's say, $y\in\Rbb^m$ is the measurement vector
derived by measurement matrix $C\in\Rbb^{m \times n}$:
\begin{equation}\label{eq:y_eq_C_Psi_s}
y = C x = C\Psi s,
\end{equation}
where $m < n$.
$C$ defines where we measure the original signal $x$,
while $y$ is the measurement result.
Eq. \ref{eq:y_eq_C_Psi_s} is saying that with $m$ points
in original signal $x$, i.e. $y$, we could still infer $s$.

Then, we find a $s$ consistent with the $y$ by solving
\begin{equation}\label{eq:l1_minimization} 
    s = \text{argmin}_{s^\prime} \| s^\prime \|_1,
    \text{ s.t. } y = C\Psi s. 
\end{equation}
Here $l_1$-minimization will yield a sparse $s$
with high probability
if $C\Psi$ satisfies
restricted isometry property (RIP).

Once we solve Eq. \ref{eq:l1_minimization}, we reconstruct
the original signal by
$x = \Psi s$.

\subsection{2-D Compressed Sensing}

Above is an example of 1-dimensional (1-D) CS.
For 2-D CS for image, general idea is to transform to 1-D CS.

Denote $n_0$ and $n_1$ as length of rows and columns separately, and
$\Psi_{n_0} \in \Rbb^{n_0\times n_0}$ as DCT basis.
Define $\vectorize(X)$ as stacking columns of matrix $X$ vertically.
If $X\in \Rbb^{n_0 \times n_1}$, then $\vectorize(X)\in\Rbb^{n_0 n_1}$.


Next, we define
$\Psi := \Psi_{n_1} \otimes \Psi_{n_0}$ and $n:=n_0 n_1$,
then we solve Eq. \ref{eq:l1_minimization} as what we do in 1-D CS.
After we obtain $x$, we reshape $x\in\Rbb^{n_0n_1}$ into $X\in\Rbb^{n_0\times n_1}$.


\section*{Acknowledgements}
We thank the anonymous reviewers for their feedback. This research was supported by the National Science Foundation award: 2212232.